**Original Paper**

**Complementary magnetic localized surface plasmons**

*Zhen Gao[1], Fei Gao[1], Youming Zhang[1], Baile Zhang[*1,2]*

*Corresponding Author: E-mail: blzhang@ntu.edu.sg (B. Zhang)

[1]Division of Physics and Applied Physics, School of Physical and Mathematical Sciences, Nanyang Technological University, Singapore 637371, Singapore.
[2]Centre for Disruptive Photonic Technologies, Nanyang Technological University, Singapore 637371, Singapore.

**Abstract**

Magnetic localized surface plasmons (LSPs) supported on metallic structures corrugated by very long and curved grooves have been recently proposed and demonstrated on an extremely thin metallic spiral structure (MSS) in the microwave regime. However, the mode profile for the magnetic LSPs was demonstrated by measuring only the electric field, not the magnetic field. Here, based on Babinet's principle, we propose a Babinet-inverted, or complementary MSS whose electric/magnetic mode profiles match the magnetic/electric mode profiles of MSS. This complementarity of mode profiles allows mapping the magnetic field distribution of magnetic LSP mode profile on MSS by measuring the electric field distribution of the corresponding mode on complementary MSS. Experiment at microwave frequencies also demonstrate the use of complementary MSS in sensing refractive-index change in the environment.

# 1. Introduction

Plasmonic metamaterials formed by texturing metallic surfaces have provided a platform of manipulating surface electromagnetic modes at the subwavelength scale ranging from microwave to far-infrared frequencies [1-3]. Since these surface waves resemble the properties of propagating surface plasmon polaritons (SPPs) at visible frequencies, they are also termed as designer, or spoof, surface plasmon polaritons [4-17]. Recently, the concept of localized surface plasmons (LSPs) in structures with closed surfaces, i.e., plasmonic particles [18], has also been proposed for this sub-visible frequency range [19-23]. Periodically textured perfect electric conductor (PEC) particles in two or three dimensions can resemble LSPs in the optical regime. Particularly, a deep-subwavelength metallic spiral structure (MSS) [24-25] can support spoof magnetic LSPs (magnetic dipole mode). However, the magnetic mode profile of previous magnetic LSPs was not directly demonstrated because only their electric mode profile was measured in experiment, while their magnetic mode profile could not be observed directly.

In this Article, motivated by Babinet's principle [26-31], we propose a complementary MSS which can also support LSP modes. We show that the electric/magnetic LSP mode profiles on the complementary MSS are exactly the corresponding magnetic/electric LSP mode profiles on the MSS. This therefore provides a unique way to map experimentally the magnetic field of magnetic LSPs by measuring the electric field of the corresponding electric LSPs on complementary MSS. Moreover, it has been shown previously that the resonance spectrum of MSS can be used to sense the refractive-index change in the surrounding environment. Here we show that the complementary MSS has better sensitivity to local refractive-index changes that are caused dominantly by dielectrics.

## 2. Magnetic localized surface plasmons

Figure 1 shows the geometry of the previously studied MSS [24-25] [Fig. 1(a)] and the geometry of the proposed complementary MSS [Fig. 1(b)]. Both MSS and complementary MSS consist of four spiral arms (or slots in the complementary structure) wrapped 1.5 turns with outer radius $R = 12.5$ mm. Each spiral arm (or slot) has a width $w = 0.5$ mm and the spacing of neighbouring arms (or slots) at the outer radius is $d = 1.5$ mm. The thickness of metallic layers in both structures is 0.018 mm.

Before comparing with complementary MSS, we first simulate the LSP mode profiles supported on MSS, with methods similar to previous studies [24-25]. Note that a 0.2 mm-thick dielectric substrate with relative permittivity $\varepsilon = 3.5$ (FR4) is adopted in all simulations to account for the practical experiment. Metal is modelled as a perfect electric conductor (PEC). Similar to previous studies on MSS [24-25], we can find two fundamental LSP mode at 1.09 and 1.85 GHz, which were termed as electric dipole mode and magnetic dipole mode, respectively. Since previous studies termed these modes based on their mode profiles [24-25], we plot in Fig. 2(a-b) the electric field distributions and in Fig. 2(c-d) the magnetic field distributions of these two modes for comparison in a transverse plane 1 mm above MSS. The electric field distribution of the first mode at 1.09 GHz is similar to that of a twisted electric dipole, and thus was termed as electric LSPs [24]. The magnetic field distribution of the second mode at 1.85 GHz is akin to that of a magnetic dipole, and thus was termed as magnetic LSPs, although it is the electric field [as in Fig. 2(b)] that was demonstrated [24]. To examine the electric and magnetic properties of these modes more clearly, we can plot the electric surface current distributions in Fig. 2(e-f), as was done in Ref. [25]. It can be seen in Fig. 2(e) that indeed the surface current of the first mode at 1.09 GHz is similar to that of a twisted electric dipole. Yet the twisting will significantly enhance local magnetic fields, and exhibit a magnetic profile as shown in Fig. 2(c). In Fig. 2(f), we can see that the surface current for the second mode at 1.85 GHz clearly form a closed loop, and thus will cause a magnetic dipole oriented out of plane.

### 3. Complementary magnetic localized surface plasmons

We then simulate the mode profiles of the complementary MSS. We can get two fundamental LSP modes at 1.06 GHz and 1.90 GHz, respectively, corresponding to the two fundamental modes previously studied on MSS. We then plot in Fig. 3(a-b) the electric field distributions and in Fig. 3(c-d) the magnetic field distributions of these two modes in a transverse plane 1 mm above complementary MSS. As predicted from Babinet's principle [26-31], the electric [Fig. 3(a-b)] and magnetic [Fig. 3(c-d)] fields on complementary MSS match the same profiles of magnetic [Fig. 2(c-d)] and electric [Fig. 2(a-b)] fields on MSS, respectively.

Before we examine these modes on complementary MSS, let us first draw a few conclusions on symmetry of their electromagnetic fields. Because all fields are driven by currents on a thin metal film, the tangential electric fields above and below the complementary MSS must be continuous ("in phase"), while the tangential magnetic fields above and below must be opposite ("out of phase"). Here, since we are mainly interested in the upper half space above the complementary MSS, we artificially delay all fields below the complementary MSS by a phase of $\pi$. This artificial phase delay will make tangential electric fields discontinuous, but tangential magnetic fields continuous, across the original metal film, and thus convert the original electric currents (originally because of the discontinuous magnetic fields) on the metal film to artificial magnetic currents (because of the discontinuous electric fields), from the viewpoint of observers in the upper half space above the complementary MSS.

We plot in Fig. 3(e-f) the artificial magnetic currents on complementary MSS for these two modes. Firstly, it can be seen in Fig. 3(e) that for the mode at 1.06 GHz, the artificial magnetic currents flow along a curved line, resembling that of a twisted magnetic dipole. Therefore, we term this mode at 1.06 GHz as magnetic LSPs. Secondly, as shown in Fig. 3(f) for the mode at 1.90 GHz, the artificial magnetic currents form a closed loop, whose fields follow an electric dipole. We thus term this mode at 1.90 GHz as electric LSPs. This duality of electric currents on MSS and magnetic currents on complementary MSS is in accordance with Babinet's principle [26].

**4. Experimental verification of complementary magnetic localized surface plasmons**

We then proceed to experimentally demonstrate the above predicted dual phenomena between MSS and complementary MSS. We fabricate MSS and complementary MSS via a standard printed circuit board fabrication process on a 0.2 mm-thick dielectric substrate (FR4) with relative permittivity $\varepsilon = 3.5$, as shown in the inset of Fig. 4(a-b). We first place a transmitting monopole antenna 1 mm away from one side of the samples to excite the surface modes, and another receiving monopole antenna at the other side of the samples to detect the resonance spectra. The locations of monopole antennas are indicated as a pair of red dots in the insets of Fig. 4(a-b). Both antennas are connected to a vector network analyzer (R&S ZVL-13). The measured near-field response spectra of MSS and complementary MSS are plotted in Fig. 4(a-b). In the measured spectra, two distinct resonance peaks are clearly observed at frequencies $f = 1.12$ GHz and $f = 1.86$ GHz for MSS [Fig. 4(a)], and $f = 1.10$ GHz and $f = 1.89$ GHz for the complementary MSS [Fig. 4(b)], respectively. The two resonances for each structure are in agreement with numerical simulation, and some minimal shifts are mainly caused by fabrication error.

Furthermore, we present the measured electric-field distributions on the MSS in Fig. 4(c-d) and the measured electric-field distributions on complementary MSS in Fig. 4(e-f), respectively. By means of a near-field scanning technique, we have mapped the local out-of-plane component of electric field (Ez) for both electric and magnetic dipole-mode LSP resonances. By comparing Fig. 4(c-d) with Fig. 2(a-b), and Fig. 4(e-f) with Fig. 3(a-b), we conclude that the experimental results match well with simulation. Interestingly, it can be seen that the electric-field mode profiles measured on the complementary MSS [Fig. 4(e-f)] follow exactly the magnetic-field mode profiles simulated on the MSS [Fig. 2(c-d)], the latter of which have never been experimentally measured before.

**5. Near-field refractive index sensing**

It should be pointed out that, although the duality between LSPs on the MSS and complementary MSS is mathematically symmetric, their responses to the same environmental change in refractive index can be different. It has been shown that [25] the LSP resonances on MSS have potential applications in refractive-index sensing, since their spectral positions depend on the dielectric environment around the structure. To demonstrate the different responses of LSP resonances on MSS and complementary MSS to the change of dielectric environment, we put a dielectric Teflon plate ($\varepsilon = 2.1$) with thickness 3 mm on both MSS and complementary MSS and then compare the resonance shifts in their near-field response spectra. The measured near-field response spectra are shown in Fig. 5(a) and Fig. 5(b), respectively. From the measured results, we observe that the resonance frequencies of both the electric and magnetic LSPs on MSS and complementary MSS experience obvious shifts after they are covered under the dielectric Teflon plate. Specifically, for the MSS [Fig. 5(a)], we obtain 0.075 GHz shift (from 1.120 to 1.045 GHz) for the electric LSPs and 0.12 GHz shift (from 1.86 to 1.74 GHz) for the magnetic LSPs, when the dielectric Teflon plate is put on top of the MSS. For the complementary MSS [Fig. 5(b)], we obtain 0.08 GHz shift (from 1.1 to 1.02 GHz) for the magnetic LSPs and 0.18 GHz shift (from 1.89 to 1.71 GHz) for the electric LSPs, when the same Teflon plate is put on top of the complementary MSS. That is to say, for the first modes of MSS and complementary MSS (electric dipole for the former and magnetic dipole for the latter), their frequency shifts have no apparent difference, while for the second modes of MSS and complementary MSS (magnetic dipole for the former and electric dipole for the latter), the frequency shift of the latter is apparently larger than the former, as can be observed by comparing Fig. 5(a) and Fig. 5(b).

The reason for the above different resonance shifts on MSS and complementary MSS lies in the duality of electric energy and magnetic energy of resonance modes on MSS and complementary MSS. Apparently, the Teflon plate covered on MSS and complementary MSS can be treated as a dielectric perturbation that mainly disturbs the electric energy of resonance modes. However, the second resonance mode of MSS is a magnetic dipole whose magnetic energy dominates, while the corresponding mode of complementary MSS is an electric dipole whose electric energy dominates. Therefore, the second resonance mode of complementary MSS should experience larger shift compared to the second resonance mode of MSS, when the same Teflon plate covers them. On the other hand, no obvious difference in frequency shift is observed comparing the first modes of MSS and complementary MSS (electric dipole for the former and magnetic dipole for the latter). This is because the electric dipole mode for MSS [Fig. 2(a)] is a twisted one, whose twisting has enhanced local magnetic fields [Fig. 2(c)]. Subsequently, the electric energy in the near field of MSS, despite being slightly larger, is not dominant compared to the magnetic energy in the near field. Therefore, as predicted from duality, the first resonance mode on complementary MSS still has significant electric energy in the near field. Therefore, we do not observe obvious difference in frequency shift for the first resonance modes on MSS and complementary MSS when the same Teflon plate covers them.

## 6. Conclusion

In summary, we have investigated the ultrathin MSS and its complementary structure by near-field microwave scanning. The duality of their near-field response spectra and their near-field mode profiles, as predicted by Babinet's principle, has been experimentally demonstrated. The magnetic-field mode profile of the magnetic LSPs on MSS, which has not been measured previously, is now obtained by measuring the electric-field mode profile of the corresponding mode on the complementary MSS. We also demonstrate that the complementary MSS is a potential candidate for near-field sensing to detect the refractive index change of the surrounding material at microwave frequencies.

**Acknowledgement**

This work was sponsored by Nanyang Technological University for NAP Start-up Grant, Singapore Ministry of Education under Grant No. Tier 1 RG27/12 and Grant No. MOE2011-T3-1-005.

Received: ()
Revised: ()
Published online: ()

**Keywords:** magnetic localized surface plasmon, Babinet's principle, spoof surface plasmon

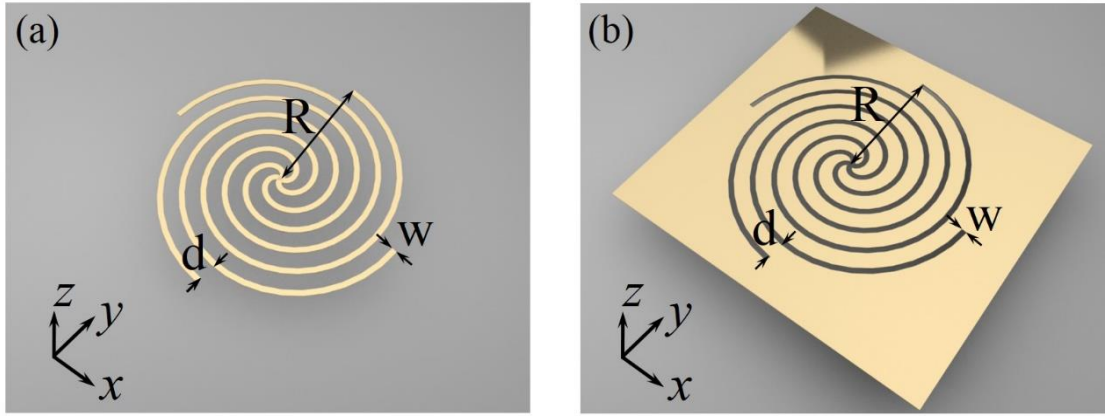

**Figure 1** Schematics of (a) the ultrathin metallic spiral structure and (b) its complementary structure. The width of the spiral arms (or slots for the complementary structure) is $w = 0.5$ mm. The spacing between neighboring arms (or slots for the complementary structure) at the outer radius is $d = 1.5$ mm. The outer radius $R = 12.5$ mm. Both structures have a thickness of 0.018 mm.

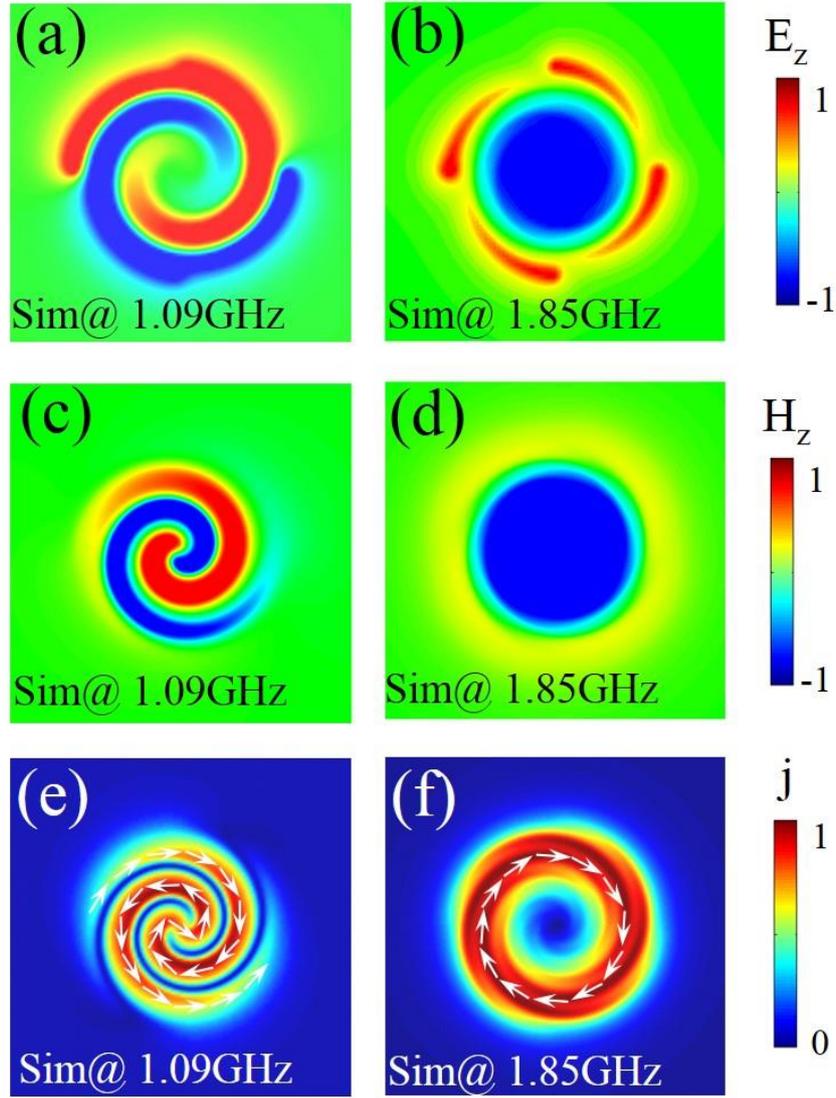

**Figure 2** Simulation of mode profiles on the ultrathin metallic spiral structure. Two fundamental modes at 1.09 GHz and 1.85 GHz are observed. (a-b) Normalized vertical (out-of-plane) component of electric field (Ez) on a transverse plane 1 mm above the metallic spiral structure. (c-d) Normalized vertical (out-of-plane) component of magnetic field (Hz) on a transverse plane 1 mm above the metallic spiral structure. (e-f) Normalized electric surface current (j) distributions on the metallic spiral structure. White arrows indicate the direction of currents.

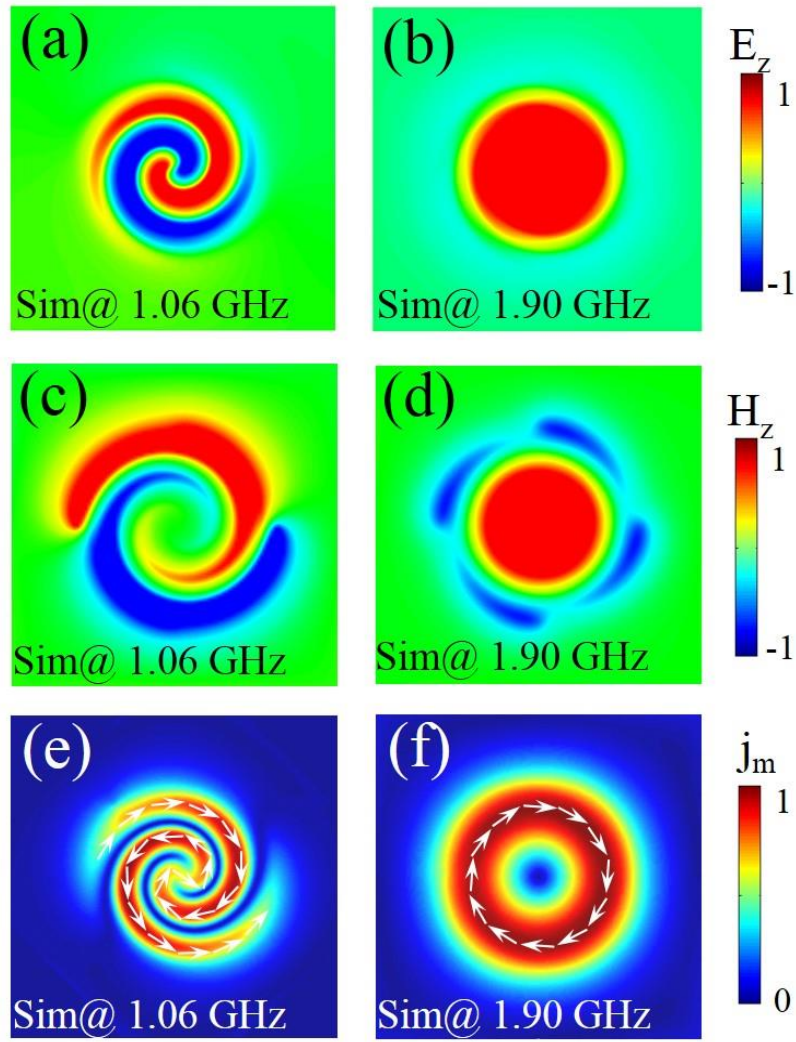

**Figure 3** Simulation of mode profiles on the complementary metallic spiral structure. Two fundamental modes at 1.06 GHz and 1.90 GHz are observed. (a-b) Normalized vertical (out-of-plane) component of electric field ($E_z$) on a transverse plane 1 mm above the complementary metallic spiral structure. (c-d) Normalized vertical (out-of-plane) component of magnetic field ($H_z$) on a transverse plane 1 mm above the complementary metallic spiral structure. (e-f) Normalized artificial magnetic surface current ($j_m$) distributions on the complementary metallic spiral structure. White arrows indicate the direction of currents.

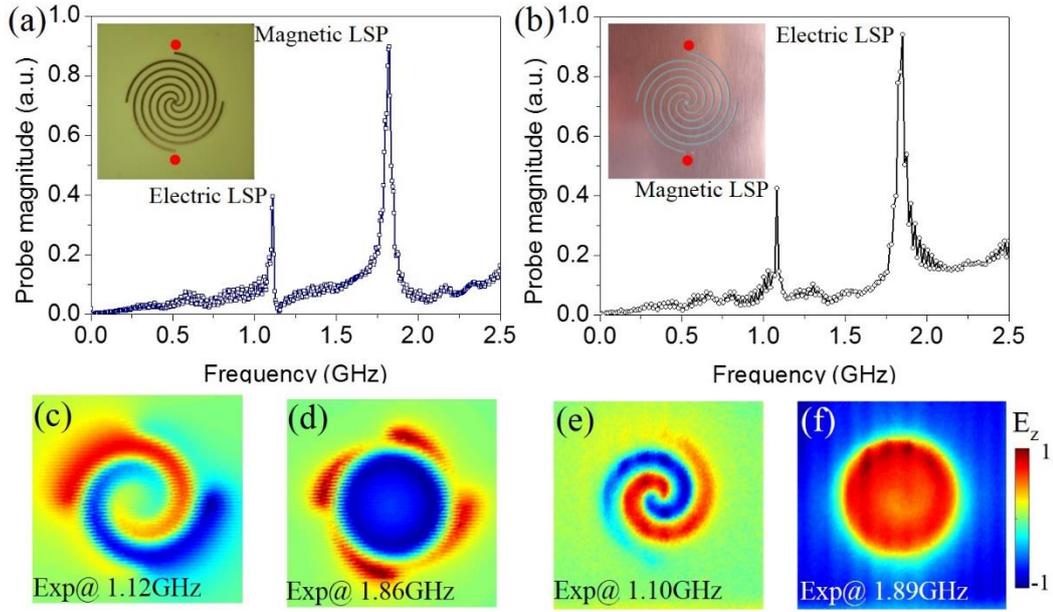

**Figure 4** Experimental demonstration of mode duality between the metallic spiral structure and its complementary structure. (a-b) Measured near-field response spectra of metallic spiral structure (a) and its complementary structure (b). Insets show photos of the fabricated samples where two red dots indicate the positions of monopole antennas in measurement. (c-d) Measured near-field patterns of normalized vertical electric field (Ez) on a transvers plane 1 mm above the metallic spiral surface at 1.12 GHz (c) and 1.86 GHz (d). (e-f) Measured near-field patterns of normalized vertical electric field (Ez) on a transvers plane 1 mm above the complementary metallic spiral surface at 1.10 GHz (e) and 1.89 GHz (f).

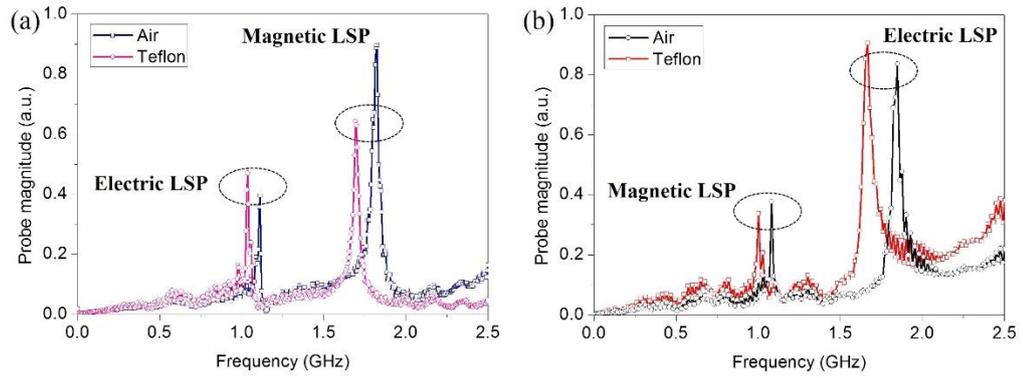

**Figure 5** Resonance frequency shifts caused by a dielectric Teflon plate. (a) The measured near-field response spectra of metallic spiral structure covered by air (blue line) and Teflon (pink line), respectively. (b) The measured near-field response spectra of complementary metallic spiral structure covered by air (black line) and Teflon (red line), respectively.

**Graphical Abstract**

Magnetic localized surface plasmons have been recently proposed and demonstrated on an extremely thin metallic spiral structure (MSS) in the microwave regime, but only their electric field was measured. Here, based on Babinet's principle, we propose a Babinet-inverted, or complementary MSS, whose electric/magnetic mode profiles match exactly the magnetic/electric mode profiles of MSS.

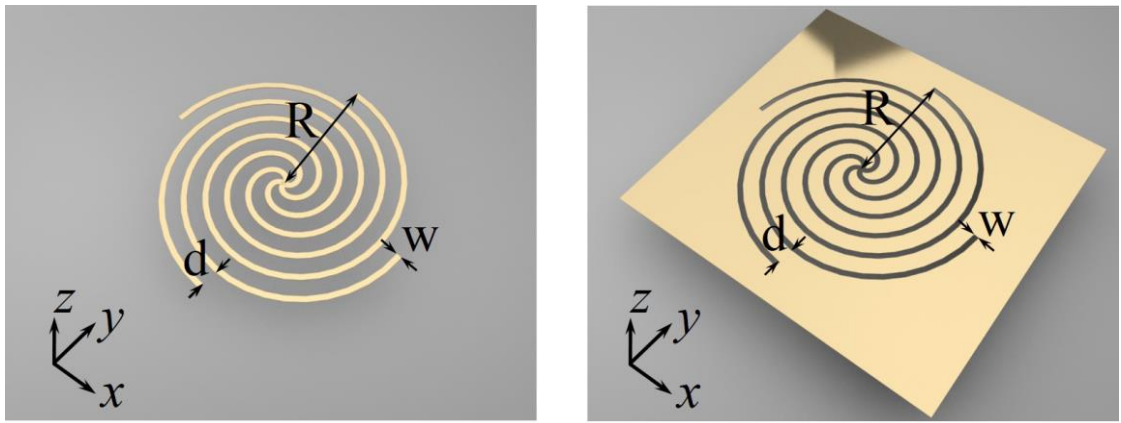